\documentclass{elsart}

\usepackage{graphicx,amssymb} 
\journal{Physica C} 

\begin{document}

\begin{frontmatter}

\title{Origin of Spontaneous Currents in a Superconductor-Ferromagnet Proximity System}

\author[Bristol]{J. F. Annett},
\ead{james.annett@bristol.ac.uk}
\author[Lublin]{M. Krawiec},
\ead{krawiec@kft.umcs.lublin.pl}
\author[Bristol]{B. L. Gy\"orffy}
\ead{b.gyorffy@bristol.ac.uk}

\address[Bristol]{H. H. Wills Physics Laboratory, 
 University of Bristol, Tyndal Ave., Bristol BS8 1TL, UK}

\address[Lublin]{Institute of Physics and Nanotechnology Center, M. Curie-Sk\l{}odowska University,
pl. M. Curie-Sklodowskiej 1, PL 20-031 Lublin,
Poland}

\begin{abstract}
We have previously shown that a ferromagnet-superconductor
heterostructure may possess a spontaneous current circulation parallel to the interface.
This current is caused by Andreev bound states in the thin ferromagnetic layer, 
and can be fully spin-polarized.  Here we investigate the total energy of the system in cases where 
the current either does or does not flow. We show that the current is a true quantum ground state 
effect, and examine the effect of the current on the different contributions to the total energy.
\end{abstract}

\begin{keyword}
proximity effect, \sep superconductor-ferromagnet interface
\PACS 72.25.-b \sep 74.50.+r, \sep 75.75.+a
\end{keyword}

\end{frontmatter}



The proximity effect between a superconductor and a ferromagnet has been a subject
 of great interest in recent years. Unlike the usual proximity effect between a superconductor (S)
 and a normal metal, a superconductor in contact with a ferromagnet (F) is subject to time reversal symmetry breaking which splits the up and down spin
 components of the spin-singlet Cooper pairs. In bulk superconducting materials a ferromagnetic exchange splitting leads to the  FFLO 
 (or LOFF) state in which the superconducting order parameter oscillates in space\cite{Fulde1964,Larkin1964}. 
 In hybrid S-F structures the same physical effect also leads to spatial oscillations\cite{Kontos2001}. This  gives rise to a  number of new phenomena, including: 
 Josephson $\pi$-junction behaviour in S-F-S  heterostructures\cite{Ryazanov2001,Frolov2004}, a giant mutual proximity effect in S-F nanostructures\cite{Petrashov1999}, and spin valve\cite{Tagirov1999} and giant magnetoresistive (GMR) effects\cite{Taddei1999}.  In addition to direct proximity effects there are there are also magnetic interaction effects between the B-field of the ferromagnet and the Meissner screening currents of the superconductor.  These lead to  the  pinning of superconducting vortices by magnetic domains
in S-F bilayers\cite{Lange2002} and by patterned arrays of ferromagnetic nanodots\cite{Lange2005}.  More complex topologial order parameter
textures have also been predicted\cite{Erdin2002}.  Lyuksyutov and Pokrovsky have recently reviewed this rapidly expanding literature\cite{Lyuksyutov2005}.

\begin{figure}[t]
 \centerline{\scalebox{0.215}{\includegraphics{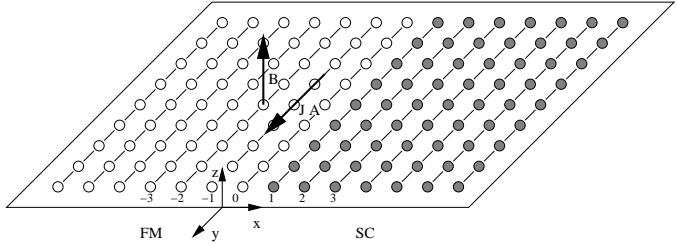}}}
 \caption{\label{Fig1} The model system we consider. The ferromagnet (FM) consists of $d$ atomic layers ($x=-d,\dots 0$),
 and is in contact with a semi-infinite superconductor ($x=1,2,\dots$). A current, $J$, may flow along $y$, parallel to the surface. This current leads to a magnetic vector potential $A$ in the same direction, and to a $B$-field in the $z$
direction perpendicular to the current, as shown. }
\end{figure}

We have recently examined the S-F proximity effect in the simple model system shown in Fig. \ref{Fig1}\cite{Krawiec2002,Krawiec2003a,Krawiec2003b}. This represents a bulk superconductor with a thin ferromagnetic layer on the surface.  By solving the Bogoliubov de Gennes equations self-consistently for this system we found that the proximity effect leads to an oscillatory
superconducting pairing amplitude
\begin{equation}
          \chi({\bf r}) \equiv  \langle c_{{\bf r}\downarrow} c_{{\bf r}\uparrow}\rangle \propto \frac{\sin{(x/\xi_F)}}{(x/\xi_F)} \label{chi-oscillating}
\end{equation}
within the ferromagnetic layer. Here $c_{{\bf r}\sigma}$ is the usual electron annihilation operator at lattice site ${\bf r}=(x,y)$ and
$\xi_F=2t/E_{ex}$ is the ferromagnetic coherence length with $t$ the hopping and $E_{ex}$ the Stoner exchange splitting within the ferromagnet. This oscillatory component has the same origin as the inhomogenous FFLO superconducting state, essentially a breaking of time reversal symmetry of the Cooper pairs, and is consistent with experiments such as those of Kontos {\it et al.}\cite{Kontos2001}.  The damping of the oscillations as a function of 
distance from the S-F interface is the same as in the usual proximity effect with non-magnetic metals, simply the decay of the Cooper pairs in the non-superconducting material.

We also found a much more surprising feature of the system shown in Fig.~\ref{Fig1}, namely a spontaneous
spin-polarized current, $J$, parallel to the S-F interface\cite{Krawiec2002,Krawiec2003a,Krawiec2003b}.  
An unexpected feature of this current is that it switches on or off suddenly as a function 
of the number of ferromagnetic layers, $d$, divided by the coherence length $\xi_F$. In fact we found the the stable 
self-consistent solution of the Bogoliubov de Gennes equations has a finite current precisely when 
the dimensionless parameter, $\Theta = 2.79 d E_{ex}/(\pi t)$ is equal to an odd integer, $n=1,3,5\dots$. 
This surprising condition precisely corresponds to the conditions for the superconducting order parameter
at the top ferromagnetic layer $\chi(-d,0)$ to be zero. As show in \cite{Krawiec2003b} the top layer
order parameter $\chi(-d,0)$  oscillates as a function of $E_{ex}$, changing sign whenever $\Theta$ is an odd integer.
Further analysis showed that this unusual condition for the current carrying solution arises from the 
spectrum of the Andreev bound states within the ferromagnetic layer. An instability arises whenever a pair of Andreev bound states become degenerate at the Fermi level.  The self-consistent solution with the current flow splits the degenerate pair, and leads to a spontaneous parallel current flow, $J$ along the S-F interface. 

In this paper we investigate in more depth this surprising sudden turning on and off of the current carrying solution.
We calculate the total energy of the system shown in Fig.~\ref{Fig1} both with and without the current flow. This allows us to explain more clearly the physical origin of the current, and to confirm that the current is indeed a true quantum ground state effect.   In the next sections we first outline the formalism and computational
methodology and  for our calculations, before we then present the results.


Our model Hamiltonian is given by the negative $U$ Hubbard model,
\begin{eqnarray}
 H =  \sum_{ij\sigma} [t_{ij} + (\varepsilon_{i\sigma} - \mu) \delta_{ij}] 
      c^+_{i\sigma} c_{j\sigma} + 
      \sum_{i\sigma} \frac{U_i}{2} \hat n_{i\sigma} \hat n_{i-\sigma},
 \label{Hamiltonian}
\end{eqnarray}
where, the nearest neighbour hopping integrals are
$t_{ij} = - t e^{-i e \int_{\vec{r}_i}^{\vec{r}_{j}} 
\vec{A}(\vec{r}) \cdot d\vec{r}}$  in the presence of a vector potential $\vec{A}(\vec{r})$. 
The site energies 
$\varepsilon_{i\sigma}$ are $0$ on the superconducting side and equal to 
$\frac{1}{2} E _{ex}\sigma$ on the ferromagnetic side, $\mu$ is the chemical 
potential, and $U_i$ is $U_S < 0$ in the superconductor and zero elsewhere.
Here $c^+_{i\sigma}$, ($c_{i\sigma}$) are the usual electron operators and $\hat n_{i\sigma} = c^+_{i\sigma} c_{i\sigma}$. 

We solve the above model in the 
Spin-Polarized-Hartree-Fock-Gorkov ($SP$-$HFG$) approximation. We work in 
the Landau gauge where $\vec{B}=(0,0,B_z(x))$ and hence $\vec{A}=(0,A_y(x),0)$ (see
Fig.\ref{Fig1}). 
Furthermore, we assume that the effective $SPHFG$ Hamiltonian is periodic in 
the direction parallel to the interface and therefore we work in $\vec{k}$ 
space in the $y$ direction but in real space in the $x$-direction.
As usual, self-consistency is assured by the relation:
\begin{equation}
 \Delta_n = U_n \sum_{k_y} 
 \langle c_{n\downarrow}(k_y) c_{n\uparrow}(k_y) \rangle = 
 - U_n \sum_{ky} \int d\omega \frac{1}{\pi} 
 {\rm Im} G^{12}_{nn}(\omega,k_y) f(\omega)
 \label{Delta}
\end{equation}
where $G^{\alpha\beta}_{nm}(\omega,k_y)$  is the $4\times 4$ Nambu Green function and $f(\omega)$ the Fermi function.

Since our model includes a  vector potential the solution may 
imply a non-zero current. For spin up electrons, in the $y$-direction this can be 
calculated from:
\begin{equation}
 J_{y\uparrow (\downarrow)}(n) = - 2 e t \sum_{k_y} sin(k_y - e A_y(n)) 
 \int d\omega \frac{1}{\pi} {\rm Im} G^{11(33)}_{nn}(\omega,k_y) f(\omega).
 \label{current}
\end{equation}
The current will give rise to a vector potential 
$A(\vec{r})$  which is obtained 
solving Ampere's law, 
$\frac{d^2 A_y(x)}{d x^2} = - 4 \pi J_y(x)$.
This full set of equations can be solved self-consistently, using the method described in \cite{Krawiec2003b}.
The various contributions to the total free energy can also be calculated, and the expressions for the
various contributing terms are given in \cite{Ketterson}.


\begin{figure}[t]
 \centerline{\scalebox{0.555}{\includegraphics{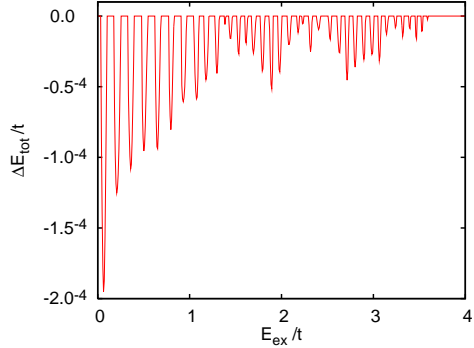}}}
 \caption{\label{Fig2} The difference in total energy between solutions with and without current flow.
 The difference is large whenever the parameter $\Theta = 2.79 d E_{ex}/(\pi t)$ is an odd integer, and 
 the current flow is stable. For other values of $\Theta$ the energy difference is zero, and no  
 current flow occurs.}
\end{figure}

The coupled set of non-linear equations given above can be solved self-consistently both
with and without a finite current flow, $J$.
Fig.~\ref{Fig2} shows the change in total energy of the system in Fig.~\ref{Fig1} between the fully
optimized self-consistent solution and the one where the current is constrained to be zero. We used the parameter
values: $d=20$ layers, temperature $T=0.01$ and $U=-2.345$ in units of the hoping integral $t$. 
Comparing the total energy of these two cases we can see, in Fig.~\ref{Fig2} that the current carrying solution is indeed the one of lower total energy, when the current carrying state is found. These are the series of sharp minima, 
each one occurring where the dimensionless parameter $\Theta$ (defined above) is an odd integer, and where the
pairing order parameter $\chi({\bf r})$ changes sign on the topmost layer of the ferromagnet.

In between these sharp minima, the energy difference goes to zero, because when $\Theta$ is away from
odd integer values the ground state has no spontaneous current flow.  One can also see in Fig.~\ref{Fig2}
that the magnitude of the energy difference gradually declines as the exchange splitting $E_{ex}$ becomes larger.
This is because the oscillatory  $\chi({\bf r})$ in the ferromagnet (Eq.~\ref{chi-oscillating}) becomes more and more heavily damped as the exchange splitting increases. The energy gain associated with the current declines, and 
for very large $E_{ex}$ the spontaneous current $J$ is suppressed.  The modulations in the size of the
energy gain shown in
Fig.~\ref{Fig2} are due to oscillations in the amplitudes of the Andreev states in the density
of states at the Fermi  energy.

\begin{figure}[t]
 \centerline{\scalebox{0.545}{\includegraphics{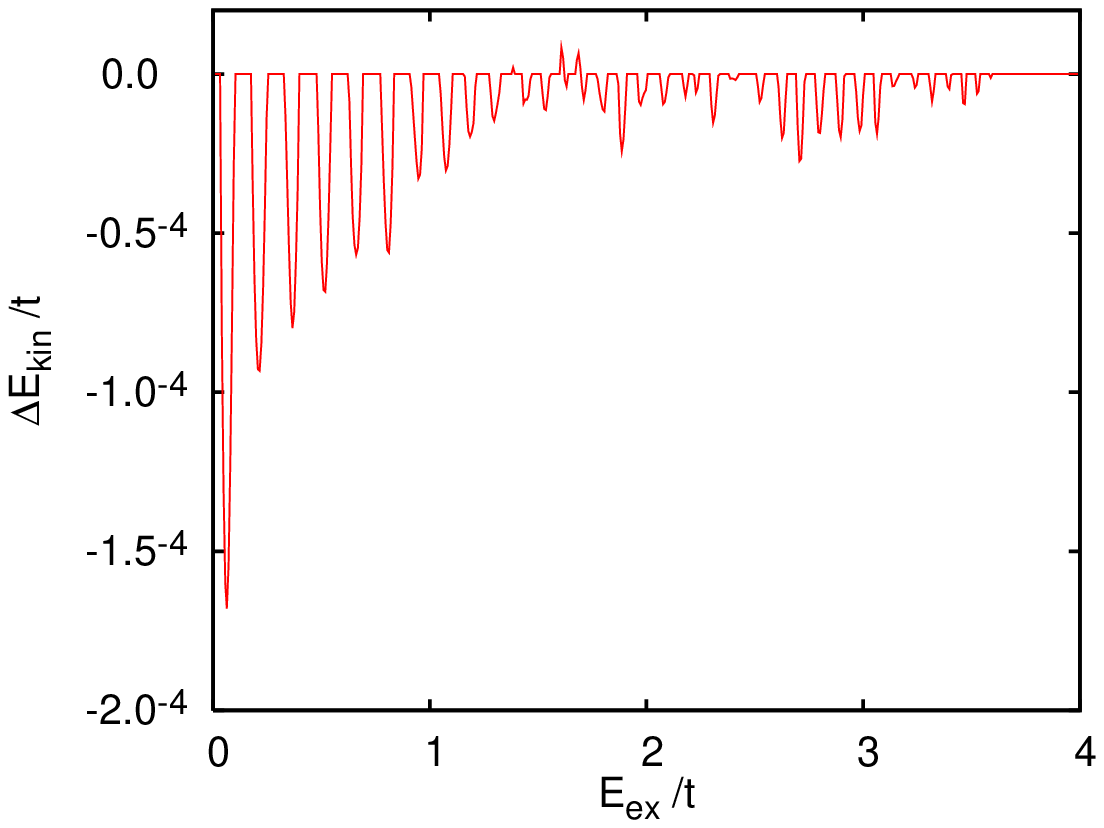}} \hspace*{0.3cm} \scalebox{0.545}{\includegraphics{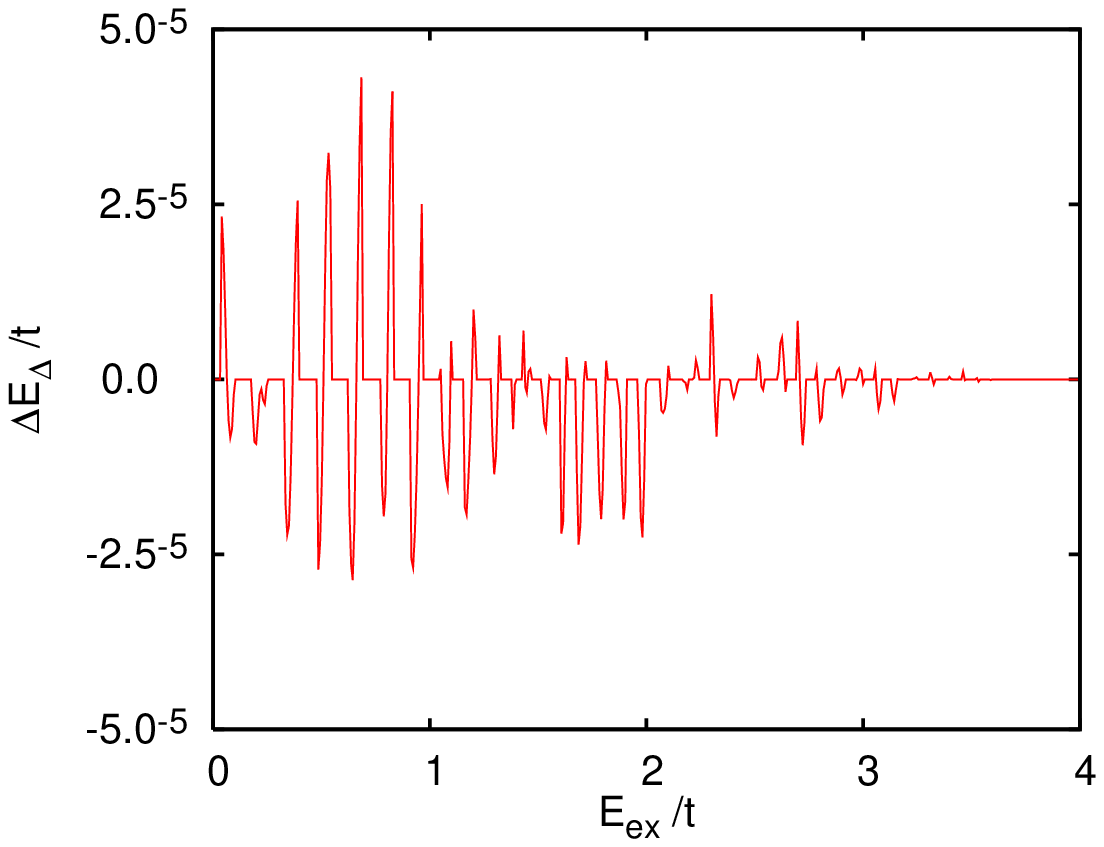}}}
 \caption{\label{Fig3} Two contributions to the total energy difference. (a) Left, shows the change in 
 electron kinetic energy $E_{kin}$. (b) Right shows the change in energy from the current term $E_J$. }
\end{figure}

One can analyze these contributions further by examining the various terms contributing to the total energy.
In Fig.~\ref{Fig3} we show the change in electron kinetic energy  $\langle t_{ij} c^+_{i\sigma} c_{j\sigma} \rangle$.
Clearly this is very similar in magnitude and form to the total energy change of Fig.~\ref{Fig2}, showing that the spontaneous current
is mainly driven by the desire to lower the electron kinetic energy.   Other contributions to the total energy change, such as the 'gap' term,
$\sum_i |\Delta_i|^2/U_i$, shown on the right in Fig.~\ref{Fig3} are not only smaller in magnitude, but can also be either positive or negative. This shows that the spontaneous current flow has only a marginal effect on these contributions to the total energy.


In conclusion we note that these results are consistent with the picture
developed in Ref.~\cite{Krawiec2003b}.  The spontaneous current  flow is generated to lift a Fermi surface degeneracy 
between Andreev bound states. 
and is primarily driven by kinetic energy optimization among the Andreev bound states. 
The conditions for this current to be observable experimentally are (i) that the ferromagnetic exchange 
$E_{ex}$ be not too large (i.e. a weak ferromagnet), (ii) the ferromagnetic thin film thickness should be thinner than the
electron mean free path, $d < l$ \cite{Krawiec2004}, and (iii) the F-S interface transparency must be greater than
about $\eta > 0.3$ \cite{Krawiec2004}.   In systems satisfying these conditions, the existence of the spontaneous current may be confirmed experimentally by observing the related magnetic flux $\sim 0.0125 \Phi_0$ per plaquette, or by STM spectroscopy on the Andreev states near to the Fermi level\cite{Krawiec2003b}.    

We thank the ESF network AQDJJ for partial support for this work.

\end{document}